# A Semi-Monolithic Detector providing intrinsic DOI-encoding and sub-200 ps CRT TOF-Capabilities for Clinical PET Applications

Florian Mueller[1], Stephan Naunheim[1], Yannick Kuhl[1], David Schug [1,2], Torsten Solf [3], and Volkmar Schulz[1,2,4,5]

[1] Department of Physics of Molecular Imaging Systems, Institute for Experimental Molecular Imaging, RWTH Aachen University, Aachen, Germany
[2] Hyperion Hybrid Imaging Systems GmbH, Aachen, Germany
[3] Philips Digital Photon Counting (PDPC), Aachen, Germany
[4] Fraunhofer Institute for Digital Medicine MEVIS, Aachen, Germany
[5] III. Institute of Physics B, RWTH Aachen University, Aachen, Germany

E-mail: {florian.mueller , volkmar.schulz}@pmi.rwth-aachen.de

**Abstract**

**Background:** Current clinical PET systems utilize detectors where the scintillator typically contains single elements of 3 – 6 mm width and about 20 mm height. While providing good time-of-flight performance, this design limits the spatial resolution and causes radial astigmatism as the depth-of-interaction (DOI) remains unknown.

**Purpose:** We propose an alternative, aiming to combine the advantages of current detectors with the DOI capabilities shown for monolithic concepts, based on semi-monolithic scintillators (slabs). Here, the optical photons spread along one dimension enabling DOI-encoding with a still small readout area beneficial for timing performance.

**Methods:** An array of 8 monolithic LYSO slabs of dimensions 3.9 x 32 x 19 mm$^3$ was read out by a 64-channel photosensor containing digital SiPMs (DPC3200-22-44, PDPC). The position estimation in the detector's monolithic and DOI direction was based on a calibration with a fan beam collimator and the machine learning technique gradient tree boosting (GTB).

**Results:** We achieved a positioning performance in terms of mean absolute error (MAE) of 1.44 mm for the monolithic direction and 2.12 mm for DOI considering a wide energy window of 300 – 700 keV. The energy resolution was determined to be 11.3%, applying a positional-dependent energy calibration.

We established both an analytical and machine-learning-based timing calibration approach and applied them for a first-photon trigger. The analytical timing calibration corrects for electronic and optical time skews leading to 240 ps coincidence resolving time (CRT) for a pair of slab-detectors. The CRT was significantly improved by utilizing GTB to predict the time difference based on specific training data and applied on top of the analytical calibration. We achieved 209 ps for the wide energy window and 198 ps for a narrow selection around the photopeak (411 – 561 keV). To maintain the detector's sensitivity, no filters were applied to the data during processing.

**Conclusion:** Overall, the semi-monolithic detector provides attractive performance characteristics. Especially, a good CRT can be achieved while introducing DOI capabilities to the detector, making the concept suitable for clinical PET scanners.

Keywords: PET, Positron Emission Tomography, Machine Learning, Gradient Tree Boosting, CRT, CTR, DOI, clinical PET, monolithic scintillator, semi-monolithic scintillator, slabs, fan beam collimator

## 1. Introduction

Positron emission tomography (PET) is a tracer-based functional imaging technique with broad application in pre-clinical and clinical domains.[1,2] Recently, devices with a long axial field-of-view, known as total-body PET, have been established, significantly increasing the system sensitivity.[3,4] The main principle relies on the radioactive $\beta^+$-decay of the used tracer, which leads via an annihilation process to two gamma photons of 511 keV energy detected by a set of radiation detectors. The radiation detector consists of a scintillation crystal converting the gamma photon into optical photons detected by a multi-channel photosensor. For the image reconstruction, the spatial position of the gamma interaction, the deposited energy, and arrival time are estimated from the detector signals.

Currently available clinical PET systems commonly employ one-layered segmented scintillation detectors coupled to the photosensor following a one-to-one or one-to-many scheme registering the active segment. These detector types typically provide a spatial resolution between 3 to 6 mm, a decent energy resolution, and good time-of-flight information.[5] For example, the current Siemens Biograph PET CT scanner utilizes crystals of 3.2 mm pitch, achieving 210 ps time resolution.[6] However, a classic single-layered radiation detector typically does not provide information about the depth-of-interaction (DOI) of the gamma interaction. This leads to parallax errors (radial astigmatism) for off-center positions and therefore to a deterioration of the image's spatial resolution after reconstruction. In the case of the Siemens Biograph system, the spatial resolution changes from 3.7 mm to 4.6 mm and 6.0 mm for radial offsets of 1 cm, 10 cm, and 20 cm for a filtered back-projection reconstruction according to the corresponding NEMA standard.[6] While minimizing parallax errors is especially important for PET scanners with a small ring diameter, e.g., pre-clinical or organ-specific devices, these numbers demonstrate the potential of DOI information for clinical scanners. Using a bench-top setup with two rotating arms of 70 cm diameter, the nearly homogenous spatial resolution along the field of view has also been demonstrated experimentally.[7] Introducing DOI is also pointed out to be beneficial for total-body PET systems: On one hand, the spatial resolution would profit in both transverse and axial direction.[8] On the other hand, the system's ring diameter could be reduced as parallax error are diminished improving the scanner sensitivity due to an increase of the overall axial acceptance angle with less detectors, and thus, reducing the costs.[9]

Different approaches have been investigated to extract DOI information from one-layered segmented scintillation detectors. For example, specific foil arrangements and surface treatments of scintillator elements were developed to introduce DOI-dependent light sharing between individual elements.[10–12] A similar effect based on light sharing was achieved by adding a uniform glass plate on top of the scintillator arrays[13,14], which was adapted to sharing light only between neighboring scintillator elements using a prismatoid light-guide array (Prism detector)[15,16]. A distinct physical effect could be utilized by partially phosphor-coating the scintillator element's surfaces, absorbing and re-emitting a fraction of the light, thus leading to a DOI-dependent signal change.[17,18]

As an alternative to segmented arrays, monolithic radiation detectors sharing the optical photons with many readout channels of the photosensor are an active area of research, as summarized in a recent overview.[19] These detectors have shown attractive performance regarding spatial, energy, and time resolution while intrinsically providing DOI information.[20–25] The spatial resolution in both planar and DOI direction depends on the calibration process and position estimation algorithm. Usually, a deterioration of the spatial resolution close to the scintillator's edges (edge effects), mainly caused by multiple reflections of optical photons, is observed. This challenge was tackled by analytical [26,27], simulated [25,28,29], and experimental [24,30–32] calibration approaches. However, complex calibration techniques and computational demanding algorithms for position estimation of the gamma interaction remain as one of the main challenges to transfer this concept into application with many detectors, such as clinical scanners or total-body PET. In most presented routines, every individual detector is calibrated with an external reference for positioning estimation which takes days up to weeks using parallel hole collimators illuminating single points. More efficient calibration setups, e.g., fan beam collimators, significantly reduce the calibration time down to hours or less which seems more feasible for scanners with many detectors.[24,30,33–35] For position estimation, a wide range of algorithms have been proposed, e.g., *k*-nearest neighbors (*k*NN) [25,36–38], maximum likelihood [31,39], Voronoi diagrams [40] or neural networks [29,41–43]. In previous work, we have established a positioning method based on the supervised machine learning algorithm gradient tree boosting (GTB), enabling an easy trade-off between positioning performance and computational requirements.[24,33,44]

Another challenge of monolithic detectors is intrinsic to their principle: Due to the light-sharing, the density of optical photons is significantly reduced compared to segmented detectors. This makes the detector more prone to the effects of the photosensor's dark counts, increases the relative error of detected optical photons per channel, and reduces high-count-rate abilities. Furthermore, all data of read out photosensor channels need to be transferred or processed, increasing the required bandwidths for the PET electronics.

Semi-monolithic scintillators, called slabs, aim to combine the advantages of both segmented and monolithic detector concepts: Optical photons are distributed along one dimension of the photosensor's readout channels, defining a monolithic and segmented direction (see Figure 1). Thus, slabs provide inherent DOI capabilities in contrast to simple segmented arrays. The density of optical photons is higher compared to large monolithic scintillators, beneficial for timing and energy resolution. Especially, the time resolution may profit compared to monoliths as it follows $1/\sqrt{n}$ with $n$ being the number of optical photons arriving on a photosensor's readout channel.[45] The dead time of the sensor gets reduced as only a fraction of readout channels is involved in single gamma interactions, additionally advantageous for band-with requirements of the PET electronics. Also, slabs are potentially cheaper than or at least competitive to large monoliths in price due to fewer rejections caused by crystal defects and geometrical aspects in the manufacturing process. We expect the price to be comparable to simple segmented arrays of 2 to 4 mm crystal pitch. As the active slab can be identified with traditional methods, the calibration effort is reduced for one dimension. Still, slabs require a calibration for gamma position estimation along the monolithic and DOI direction and thus share the related challenges of monolithic detectors.

To the author's knowledge, the concept of slabs was first presented in the field of single photon gamma cameras for the SPRINT devices dedicated for head imaging in the 1980s.[46,47] A simulation study in 2008 [48] revisited the concept for PET imaging and was later experimentally validated using 2 mm thick and 10 mm high slices coupled to a PMT array [49]. Further research demonstrated detectors using slabs of 1 mm thickness in segmented direction and 10 mm height.[50] These slabs required an additional lightguide to share optical photons along the segmented direction as their pitch was smaller than the readout channels of the photosensor. Later work increased the length along the monolithic direction to reduce the fraction of crystal volume affected by edge effects.[51] Most recently, Zhang et al. presented an adaption of their detector for clinical applications increasing the height to 20 mm and thickness to 1.37 mm while still keeping a lightguide to resolve the slabs along segmented direction.[52] The detector utilized a SiPM array of 3 mm channel pitch and was read out by the TOFPET2 ASIC (PETsys electronics). A spatial resolution of 2.43 mm FWHM along the monolithic direction and 4.77 mm FWHM for DOI was achieved for one-sided readout. The energy resolution was reported as 25.7 % and the time resolution as 779 ps. An alternative approach employing the idea of slabs was followed to establish a trapezoidal detector for high-resolution imaging with a maximal packing fraction of scintillator material in the PET gantry, presented both in simulation and experimental work.[53,54] The experimental work studied slabs of 0.65-0.93 x 28.8 x 8 mm$^3$ coupled with mirror film of varying dimensions within the array to control light sharing along the segmented direction.[54] The latest update of this slab-concept studied different dimensions increasing the thickness up to 10 mm height while simplifying the manufacturing process of the slab arrays.[55] Another detector utilizing the TOFPET2 ASIC of an array of 24 slabs of 25.8 x 12 x 0.95 mm$^3$, suitable for high-resolution imaging, was examined with a special focus on timing performance.[56] Including all developed optimization and corrections, i.e., time-walk correction and weighting of multiple timestamps, a time resolution of 275 ps was reported for an energy window of 410 – 610 keV.

In this work, we introduce the concept of a slab-based detector targeting the needs of clinical whole-body or total-body PET systems. The detector consists of an array of 8 slabs of 4 mm pitch and 19 mm height. The scintillator pitch was chosen to follow a one-to-one coupling scheme along the segmented direction with the utilized digital SiPM (DPC 3200-22-44, Philips Digital Photon Counting, PDPC); thus, no lightguide is needed. The photosensor contains 64 readout channels grouped in 16 independent trigger regions. We present a full characterization regarding spatial resolution in monolithic and DOI direction, energy resolution as well as time resolution. All position calibrations employ a fan-beam collimator setup for time-efficient irradiations and GTB as position estimation algorithm. For energy and timing calibrations, coincidence flood irradiations without collimator were used. We established both an analytical and machine-learning-based timing calibration approach. All developed algorithms deal with missing information regarding single gamma interactions originating from the independent trigger regions, such as non-available readout channel information, to avoid losing sensitivity by rejecting data.

## 2. Materials

### 2.1. Radiation Detectors

#### 2.1.1. Photosensor

In this work, we employed an array, usually referred to as tile, of 16 digital SiPM (DPC 3200-22) arranged in a 4 x 4 matrix of 32.6 x 32.6 mm$^2$ outer dimensions (PDPC).[57,58] The DPCs provide 2 x 2 readout channels (pixels) with 3200 SPADs each on 3.2 x 3.88 mm$^2$ active area. Every single SPAD is integrated into an individual logic circuit for charging, readout, and selective disabling of the SPAD. Utilizing a dark count measurement, 10 % of the noisiest SPADs were deactivated to reduce the overall dark count rate.[59] Each individual DPC applies a two-level trigger scheme to register and validate optical photon flashes generated by gamma particle interactions. Every pixel of the DPC is divided up into four sub-pixels which builds the basis of the DPC trigger network. Different Boolean interconnections of the sub-pixels realize four configurable trigger schemes. For example, connecting all four sub-pixels via OR gates, a first photon trigger is realized (trigger scheme 1). For trigger scheme 4, all sub-pixels are ANDed leading to a higher number of required optical photons to generate the trigger. More detailed information on required optical photon thresholds can be found in.[60] Typically, trigger schemes requiring fewer optical photons lead to better coincidence resolution time (CRT) [61] and thus, trigger scheme 1 was used in this study.

After a trigger is generated, the DPCs expect a higher number of detected optical photons to achieve the validation threshold within a set validation time. We set the validation threshold to validation scheme 16 (0x55:AND) requiring on average 54±19 optical photons. In case the validation criterion is met as well, the integration time starts leading to the final readout phase, transmission of the digital information towards an FPGA and a recharging cycle. The DPC generates a hit and sends out the photon counts of all four pixel counts and a timestamp. In the case the validation is not successful, the trigger is dismissed and the DPC activates a fast recharge.

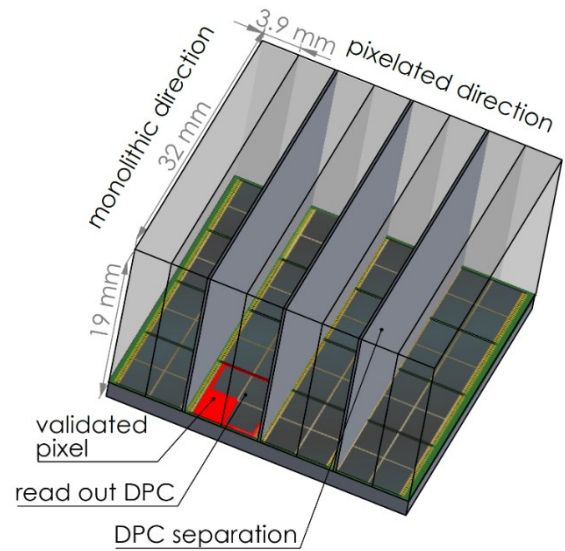

Figure 1 Setup of the slab detector stack. An array of 8 monolithic slabs with all surfaces polished is optically coupled to the photosensor. Every slab is coupled to a row of pixels. Always two slabs are separated with specular reflective foil to reduce light sharing between rows of DPCs. The specular reflective foil of all outer surfaces is not shown.

As described, every DPC is independent. Thus, not all available DPCs generate hits, especially for low optical-photon densities, leading to potentially unstable sets of present readout channel information for the gamma photon interactions. The tile offers a neighbor logic feature to force a readout of the whole tile after one DPC has validated.[59,62] However, neighbor logic is not applied as slabs not included in the gamma photon interaction would undergo a dead time and, thus, reduce the detector's sensitivity.

### 2.1.2. Slab Scintillator Configuration

We studied an array of 8 monolithic LYSO slabs (Crystal Photonics, Sanford, Florida, USA) of dimensions 32 x 3.9 x 19 mm$^3$ (see Fig. 1) optically coupled to the photosensor with Meltmount (Cargille Laboratories, Cedar Grove, NJ, USA). The LSYO slabs were polished on all surfaces. The scintillator arrangement was matched to the readout characteristics of the photosensor: Every single slab was positioned on a photosensor's pixel row in a one-to-one configuration; a pair of slabs covered the area of a row of DPCs. A specular reflective foil (ESR foil) separated pairs of slabs to reduce light sharing between DPC rows; no further optical coupling was inserted between the individual slabs. Thus, the readout area of the photosensor per gamma interaction shall be kept small. All four DPCs covered by a slab generate a hit in the optimum case. Furthermore, no active area of the photosensor was covered as the specular reflective foil was placed over the bond wires located between the DPC rows. ESR foil was attached to all outer surfaces of the scintillator array using a thin film of optical transparent adhesive tape.

### 2.1.3. 1-to-1-Coupled Coincidence Detectors

As coincidence detector, we optically coupled an array of 8 x 8 LYSO elements of 3.9 x 3.9 mm$^2$ base area and 19 mm height with a pitch of 4 mm to the same photosensor type in a one-to-one configuration. All scintillator elements were separated with ESR foil. Furthermore, ESR foil was applied to all outer surfaces as well. A pair of one-to-one-coupled detectors was prepared for reference CRT measurements.

## 2.2. Coincidence Calibration Setup

### 2.2.1. Collimator Setup

A fan beam collimator setup was used for the planar and DOI positioning calibration. The setup consists of two $^{22}$Na sources and an electrically driven translation stage (LIMES 90, Owis, Staufen im Breisgau, Germany) to irradiate the detector under calibration at known positions by moving through a fan beam.[34] At both sides of the collimator setup, the beam width can be changed independently by adjusting the corresponding slit width. The slit width for the coincidence detector was chosen to be bigger compared to the calibration detector slit to avoid losing coincidences due to geometrical aspects. The slit width was set to 0.5 mm for all positioning estimation calibration steps for the calibration detector.

The complete setup was located inside a climate chamber to ensure a light-protected environment and stable measurement conditions regarding temperature and humidity.

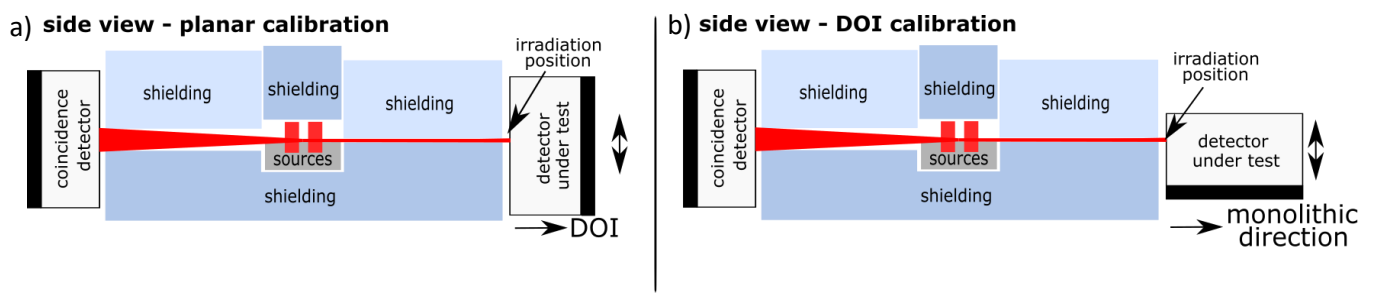

*Figure 2 Overview of the positioning calibration routine. The detector under test is stepped through the gamma beam shaped by the fan beam collimator. The irradiation position was used as the reference for training and testing the GTB models. a) Calibration of the monolithic direction. b) Calibration of DOI. Figure adapted from Müller et al., 2019, CC 3.0.*

### 2.2.2. Flood Irradiation Setup

The flood irradiation setup used for energy and analytical timing calibration was equipped with a single $^{22}$Na sources (see Figure 3a). In between the coincidence and calibration detector a source holder was located, which shifts the source along all room axes using a programmable translation stage system.[63] The two detectors were mounted facing each other at a fixed distance of 435 mm (see Figure 3b). Likewise the collimator setup, the flood irradiation setup was located inside a climate chamber for light-protection and stability of environmental conditions.

## 3. Methods

### 3.1. Data Acquisition and Preprocessing

#### 3.1.1. Positioning Calibration

Data, which was used for the positioning calibration, was acquired by measurements conducted with the collimator setup utilizing different detector orientations (see Figure 2). The activity of the $^{22}$Na sources was 12 MBq each. Both measurements were performed with a stable DPC temperature of 5 °C.

For planar positioning calibration, the detector under calibration was irradiated perpendicular to its top side, while the monolithic slab axis was orientated in a way, such that the fan beam irradiates every slab simultaneously. The crystal was moved into the beam in steps of 0.25 mm. An irradiation time of 1,000 s per position was chosen while the required number of gamma interactions used later was recorded within approximately 40 s. Thus, a total measurement time of below 2 h is feasible.

Regarding DOI positioning calibration, the detector was rotated with the aim of side irradiation of the calibration crystal at known DOI-positions. The crystal was moved into the beam in steps of 0.25 mm with an irradiation time of 500 s per position. The equivalent required irradiation time was found to be approximately 90 s for the later used gamma interactions enabling a calibration within less than 2 h.

#### 3.1.2. Energy and Analytical Timing Calibration

To acquire data for the energy and analytical timing calibration, a flood irradiation setup was used, which was equipped with a 12 MBq $^{22}$Na source. A grid of 25x25 radiation source positions with a pitch of 1 mm around the detector's geometrical center was measured in a plane parallel to the surfaces located exactly in the middle between the two detectors. At each grid position, the measurement time was set to 300 s. During the measurement, the slab detector had a temperature of 5 °C and the 1-to-1-coupled detector of 2 °C. This slight temperature difference was observed as both sensors share the same air-cooling unit and the segmented detector consumes less energy due to fewer triggering. While the energy calibration utilized all available data, the analytical timing calibration employed only a smaller subset regarding the grid positions comprising 81 radiation source positions.

#### 3.1.3. Machine Learning-based Timing Calibration

Data used for the machine learning-based timing calibration was recorded similarly to the analytical timing calibration. This time, the 12 MBq $^{22}$Na source was stepped through a grid comprising 5x5 positions with a pitch of 6 mm around the detector's geometrical center position within a plane parallel to the detectors surfaces, and additionally the plane was moved in 5 mm steps between the detectors giving in total 47x5x5 measured source positions. At each grid position, a measurement time of 120 s was used, resulting in a total measurement time of approximately 1.6 days. The temperatures were set as in the previous measurements.

#### 3.1.4. Preprocessing and Data Set Generation

We applied the same preprocessing scheme similar as described in Schug et al. 2015 for all measured data: First, the single photon counts were corrected for saturation effects using a simple logistic saturation model. Then, all DPC hits related to one gamma interaction were merged to clusters using a cluster window of 40 ns. The clusters were not restricted to the region of single slabs to be sensitive for scatter between slabs. Afterward, we selected coincidence events with a coincidence time window of 10 ns based on the first timestamp of the cluster. Clusters with less than 400 optical photons were discarded to remove noise. This threshold compares to an uncalibrated 511 keV photopeak found for 2300 optical photons. Clusters passing the coincidence and photon count criterion are called events.

For planar and DOI positioning, the events recorded utilizing the fan beam collimator setup were divided up into training, validation, and test data. The planar positioning data set contained 10,000 events per position for training, 2,500 for validation and 2,500 for testing randomly selected from the perpendicular irradiation of the top surface. For the side irradiation, we expect an exponential distribution of γ-particle interactions along the monolithic axis following to Lambert-Beer's law. Since the calibration data was used in the following steps to train and evaluate a machine learning-based positioning model, an exponential behavior would introduce a bias in the DOI positioning training. Thus, the side irradiation data was prepared to provide a constant gamma interaction density along the monolithic direction with the assistance of already established planar positioning models. Then, we randomly selected 10000 events per position for training, 2,500 for validation and 2,500 for testing.

The calibration data set for energy and timing relied on the flood irradiations and comprised 12.5 million events. For energy calibration, the data were equipped with the 3D-estimated interaction position. For timing calibration, the energy estimate was added as well.

### 3.2. Position Estimation

#### 3.2.1. Positioning Performance Parameters

All positioning performance parameters are based on the positioning error distribution of the test data sets (irradiation position $Y_i$ – estimated position $\widehat{Y}_i$). If one single parameter is given, it was calculated on the base of all available test data without distinguishing between the irradiation positions or single slabs. As the positioning error distribution shows a complex behavior, we present multiple positioning performance parameters probing different aspects of the distribution:

1. *Mean Absolute Error (MAE):* The mean absolute positioning error $1/n \sum_i |Y_i - \widehat{Y}_i|$.

2. *Root Mean Squared Error (RMSE):* The root mean squared error of the positioning error $\sqrt{1/n \sum_i (Y_i - \widehat{Y}_i)^2}$. This parameter is used as loss function during training of the GTB models, as explained below.

3. *Spatial Resolution (SR):* The FWHM of the positioning error distribution. We determine the SR following the NEMA NU 4-2008 procedure.[65] As this process includes more uncertainties due to the applied binning and fitting process, no direct comparisons of these values are given within this work.

4. *Percentile Distance dx:* The distance encloses the given percentile $x$ of the absolute positioning error distribution.

5. *Bias vector:* The mean positioning error for a given irradiation position. The bias vector is usually observed for positions close to the edge of the detector, behaving symmetrically around the crystal center. Thus, the value does not follow a Gaussian distribution and the mean usually cancels out for averaging all irradiation positions. Therefore, we report the 90[th] percentile of all individual irradiation positions' absolute bias vector values (bias90).

While MAE and RMSE are influenced by the whole positioning error distribution, SR as well as d50 focus on the central region, and d90 represents the tail region. The bias vector influences all parameters for an evaluation of the whole detector. All positioning performance parameters are not corrected for the finite beam width of 0.6 mm of the collimator setup.

### *3.2.2. Slab Identification*

For positioning an event in pixelated direction, we identified the pixel with the highest photon count and matched its geometrical position to the coupled scintillator slab. In case multiple slabs showed a pixel with the same photon count, the slab with the highest total number of optical photons was selected.

### *3.2.3. Gradient Tree Boosting Position Estimation*

The GTB algorithm for both planar and DOI position estimation is presented in detail in previous work.[24,33] These papers describe protocols for optimizing GTB models regarding both best positioning performance and computational requirements. Thus, only the main characteristics of the GTB algorithm, it's hyperparameters and optimization techniques are recapitulated here.

GTB, classified as a supervised machine learning technique, employs training data with known irradiation positions to build predictive regression models. The implementation handles partially missing data (i.e., not available DPC hits) and arbitrary input features.[66] During training, the GTB algorithm establishes an ensemble of sequential binary decisions (decision trees) which are evaluated as simple comparisons with two possible outcomes.[67,68] The training scheme follows an additive manner: The first decision tree is trained on the known irradiation position. Every following decision tree acts on the residual of the already existing ensemble of decision trees (known irradiation position – predicted position). The final predicted position equals the sum of all individual predictions of the ensemble. Thus, the evaluation of a trained GTB model can be done fully in parallel in contrast to the training process. For the objective function, we employed the RMSE.

Four hyperparameters are of particular importance in this work:

1. *Number of decision trees of the ensemble*
2. *Maximum depth:* The maximum number of binary comparisons within an individual decision tree. The decision trees are not forced to be fully grown.
3. *Learning rate:* The learning is a constant factor ranging from (0,1] applied to the residuals before training a further decision tree. High learning rates reduce the number of required decision trees to reach the local optimum positioning performance while the positioning performance decreases compared to models trained with smaller learning rates.
4. *Input features:* As shown in previous work, adding pre-calculated features carrying additional information of the (physical) problem, e.g., the first moment of the light distribution, significantly improves the positioning performance of GTB models. This allows to build models of fewer decision trees and maximum depth for a given learning rate.

Trained GTB models can be evaluated in multiple architectures, e.g., CPUs, GPUs and FPGAs.[69,70] Recently, we have demonstrated a high-performance CPU implementation of GTB in a dedicated online processing platform for the Hyperion II$^D$-scanner.[44] The study showed a clear advantage regarding data throughput in case the complete GTB model fits into the cache of the used CPU. Another approach, significantly reducing the amount of data that needs to be sent out of the system, is an evaluation of the GTB models close to the detector in an FPGA, as principally demonstrated for the successor of the Hyperion II$^D$ system architecture.[71,72] As the resources of typical FPGAs remain limited, especially regarding available memory in case no additional RAM is added, the model complexity of the trained GTB models needs to be restricted. The model complexity can be translated into the memory requirement (MR) of the ensemble as an objective measure. For a single decision tree, the MR can be estimated by

$$\text{MR}(d) = (2^d - 1) * 11\,\text{B} + 2^d * 6\,\text{B}$$

with $d$ the maximum depth.[24] As the MR represents a simply comprehensible parameter relevant for both CPU and FPGA implementations, we will use the MR to compare and discuss the complexity of trained GTB models.

### 3.2.3.1.   General Optimization Protocol

Protocols to study the general characteristics of GTB models for both planar and DOI positioning were presented in previous work.[24,33] Here, both planar and DOI positioning share the same settings if not stated otherwise. Based on previously obtained results, we selected the following hyperparameter ranges as a starting point for a full grid search: The maximum ensemble size was set to 1000 decision trees, the maximum depth was ranged between 4 and 12 in steps of 2, and the learning rate was chosen from {0.05, 0.1, 0.2, 0.4, 0.7}. The number of training events per irradiation position was fixed to 10,000 for planar positioning and DOI positioning, respectively, with a pitch of 0.25 mm for all datasets.

As stated before, additional calculated features typically improve the positioning performance of GTB models. The calculated features included the index numbers of the hottest pixel and DPC, first and second moment of the light distribution, the total sum of detected optical photons and a projection of the photon counts perpendicular to the slabs. Furthermore, the sum of squared pixel intensities, which has proven as a useful measure for DOI, was added.[33] All calculated features were defined for the complete photosensor array without interpolating for missing DPC hits and not constrained to the active slabs. During the calculation of the additional features, we artificially set photon counts corresponding to missing DPC hits to 0. In the raw data, the GTB models can distinguish between a zero-photon count and a missing DPC hit as they are masked with a negative value. Previous work revealed the influence of adding the estimated planar interaction position to train and evaluate DOI models to be smaller than 0.7 % for the best-achieved positioning performance. As this would require sequential processing for planar and DOI position estimation in the PET system architecture, this option was not used.

### 3.2.3.2.   High-Positioning-Performance Optimization

We selected the GTB models with the best MAE based on all available validation data without applying any restrictions regarding memory requirement except the boundary condition of the maximum 1000 decision trees. The MAE was chosen as it is influenced by all parts of the positioning error distribution. The positioning performance parameters including the positioning error distribution, cumulative positioning error distributions as well as spatial distribution of the MAE were then calculated for the fully unseen test data set. All parameters are presented without any energy window (only the minimum required number of 400 optical photons), for a wide energy window of [300 keV, 700 keV] and a narrow window of [411 keV, 561 keV]. Nonetheless, training and validation of the GTB models were performed without any filter as a correct positioning is required to calculate the energy as explained below.

### 3.2.3.3.   Memory-Requirement Optimization

To study the trade-off between complexity of the GTB models represented as MR and the positioning performance, we employed the concept of Pareto efficiency. The Pareto efficiency describes the situation where no goal can be improved without worsening the other one.[73] Therefore, we plotted all trained GTB-models as individual points and constructed the Pareto frontier (see Figure 7). The method was identical for both planar and DOI position estimation. In previous studies[24,33], we selected the most efficient models based on a specified gradient of the positioning performance. The Pareto frontier model provides the advantage that no pre-selection and manual definitions are included. Using the Pareto efficiency, the optimal GTB models for implementations with the aspect of high data throughput and online processing can be selected.

### 3.3. Energy Calibration and Estimation

#### 3.3.1. Energy Calibration

We established a 3D-dependent energy calibration based on the estimated interaction position available through the GTB models. The detector volume was divided up into 8 voxels of equal size along the pixelated direction and 4 for the DOI direction. We determined the 511-keV-photopeak, and thus the voxel conversion factor. For the calibration, we selected all events fulfilling a neighborhood criterion requiring all direct neighboring DPCs of the hottest DPC to be read out for the active slab. This requires 3 out of 4 present DPCs for events with the hottest DPC located in the middle and two DPCs for those at the edges. The criterion was chosen based on the observed readout characteristics as shown in Figure 4, stating 95% of all events having at least 3 DPCs hits present. Though, the readout characteristic does not account for the spatial distribution of read out DPCs. Thus, the rate of all events fulfilling the DiesFN-criterion is expected to be less and turned out to be 87% of all events as reported in more detail in Section 4.3. The photon sum of all events fulfilling the DiesFN criterion for the hit slab were filled in a histogram to determine the photopeak position taking only the corresponding DPCs into account. These light distributions were then normalized and averaged to create one mean event per voxel. The resulting mean events were the base to interpolate for missing DPC hits. In case the number of events was less than 1,500 for a voxel, no calibration was performed. As stated before, the photon counts were corrected for saturation effects during the data preprocessing.

#### 3.3.2. Energy Estimation

For estimating the energy of an event, the event was first sorted into its corresponding scintillator voxel. We distinguished three cases:

a) *Events fulfilling the DiesFN criterion:* In case all photon counts are available, the detected optical photons were summed and the energy calculated using the voxel conversion factor.

b) *Events not fulfilling the DiesFN criterion:* In case of missing photon counts, the fraction of the mean light distribution corresponding to the missing pixels was estimated using the voxel's mean event. Afterward, the sum of the detected optical photons was calculated and scaled up to calculate the final energy using the voxel's conversion factor.

c) *Events without a voxel calibration:* In case an event is sorted into a voxel with no calibration present, the geometrical closest voxel was selected to continue with case a) or b). The replacing calibration was chosen randomly if multiple voxels had the same geometrical distance.

We report the energy resolution as the FWHM of the 511 keV photopeak for all available events, those fulfilling the neighborhood criterion, and interpolated ones for all voxels. The results are compared with the width of the uncalibrated distribution of the total sum of detected optical photons.

### 3.4. Timing Calibration and Estimation

The timing calibration aims to correct time skews originating from the electronic components of the detector (e.g., different signal trace length of the PCB or firmware-related differences) and time skews resulting from crystal-related domains (e.g., differences in the optical coupling, light transport, and light collection efficiencies). Two different calibration techniques were established, namely an analytical timing calibration and a machine learning-based timing calibration.

### 3.4.1. Analytical Timing Calibration

The analytical calibration process comprised concatenated sub-calibrations (in the following referred to as stages), where at each stage a specific kind of time skew is addressed. The same mathematical principle was used within each stage, relying on the formulation of a matrix equation, which was subsequently minimized. The calibration data was obtained by measuring coincidences using the flood irradiation setup. A detailed mathematical description of the process is given in Appendix I. Each of the two facing detectors is separated into abstract objects, which can be represented by physical readout channels (like pixels) but also by theoretical constructs (like scintillator volumes). Time skews are estimated within an arbitrary calibration stage by sorting events based on a defined criterion (e.g., hottest DPC, first DPC) into the corresponding channel pairs. For each possible channel pair $(i, j)$, a set of $n_{i,j}$ time differences are obtained, which generate a Gaussian distribution.

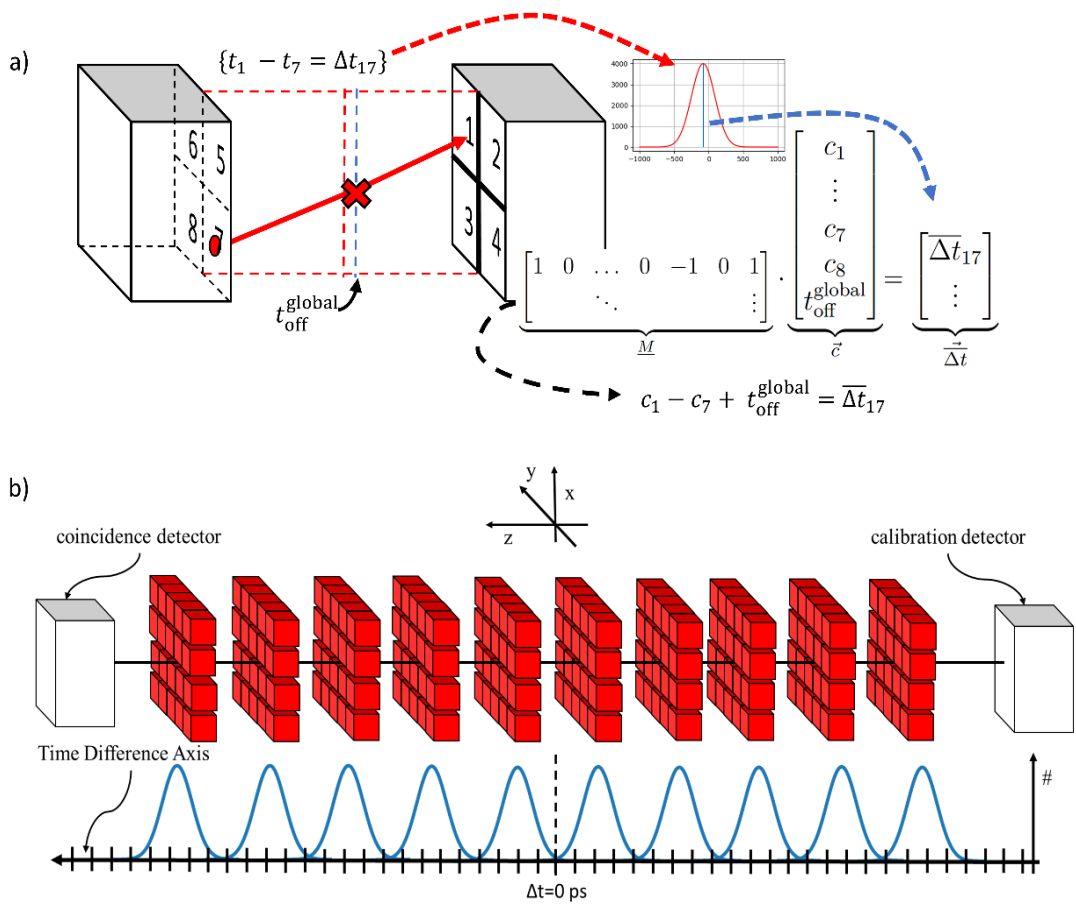

*Figure 3 Principle of both timing calibration schemes. a) Analytical timing calibration. The analytical timing calibration contains several subsequent calibration stages targeting time-skews of different origins, e.g., electrical and optical time skews. The given detector pair with four channels each is divided into different sub-volumes. For every channel combination (here channel 1 and 7), the time difference histogram is created and the mean value determined using a Gaussian fit. The mean value is inserted as one equation. After iterating over all channels pairs, the overdetermined matrix system is solved. b) Machine-learning-based calibration. The machine-learning-based calibration applies the already established correction found by the analytical timing calibration and corrects for further effects. A source is shifted between the detector pairs to generate data with known off-centred positions. To include all valid lines of responses, the source is moved in the shape of an array for every measurement plane. The expected theoretical time difference can be calculated and used as label for the measured data to train the GTB-based correction model.*

The fitted means of these distributions are used to formulate a matrix equation, while the channel combinations are encoded as a product of a matrix $\underline{M}$ and the channel vector $\vec{c}$, and the mean time differences are used to fill a mean time difference vector $\overrightarrow{\Delta t}$. Subsequently, the system is minimized using:

$$\overrightarrow{c_{min}} = \underset{\vec{c}}{\mathrm{argmin}} \, || \overrightarrow{\Delta t} - \underline{M} \cdot \vec{c} \, ||_2.$$

This procedure was repeated several times using different kinds of channels every time. For the presented results, the analytical timing calibration comprises six different calibration stages. The first two stages utilized the DPC sensors and pixels as channels, and the following stages were based on changing voxel volumes.

### 3.4.2. Machine Learning-Based Timing Calibration

The machine learning-based timing calibration was applied on top of the previously performed analytical timing calibration and uses the already corrected timestamps, including the first stages of the analytical timing calibration. To generate data for training, validation and testing, the radiation source was shifted between the facing detectors to known off-centered positions. Based on the known radiation source position, the expected theoretical time difference can be calculated and used as label for the measured data. Like the positioning procedure (see Section 3.1.1), GTB was used as supervised machine learning algorithm. We included the difference of the first timestamps of both detectors, the first 4 timestamps of both detectors, the estimated interaction position as well as photon counts corresponding to the included timestamps as features for the model training. As for the position estimation, a grid search was conducted to find the best performing GTB model. The maximum depth was restricted to {4, 12, 20}, the learning rate to {0.1, 0.3, 0.5, 0.7}, and the number of decision trees to 300. The training was stopped as soon as overfitting was detected.

### 3.4.3. CRT Estimation & Evaluation

The CRT is based on the FWHM of the time difference distribution between the first timestamps of coincidence and the calibration detector. Considering that the generated distribution may be slightly skewed due to a combination of different crystal geometries (e.g., slab detector and one-to-one coupled detector), a numeric FWHM estimation is performed instead of a naive Gaussian fit. For the general case of two detectors "a" and "b", the CRT of a mixed system "$ab$" can be extrapolated to a pure system "$aa$" if the timing performance of the opposite pure system "$bb$" is known,

$$\mathrm{CRT}_{aa} = \sqrt{2 \left( \mathrm{CRT}_{ab}^2 - \frac{\mathrm{CRT}_{bb}^2}{2} \right)}.$$

All here presented time resolution evaluations were based on unseen data, meaning that an analytical and/ or machine learning-based calibration model was built using calibration data and then subsequently applied to unseen evaluation data to report an unbiased CRT estimation. The evaluation was performed for the same energy windows as defined for the study on the positioning performance. To study the best possible CRT, we introduced an additional filter F on the light distribution ensuring the first timestamp originating from the hottest DPC sensor, the hottest pixel being on the hottest DPC sensor, and the active slab completely read out. Filter F was not included in any further processing except stated otherwise as the sensitivity of the detector was not be compromised.

## 4. Results

### 4.1. General Detector Readout-Characteristics

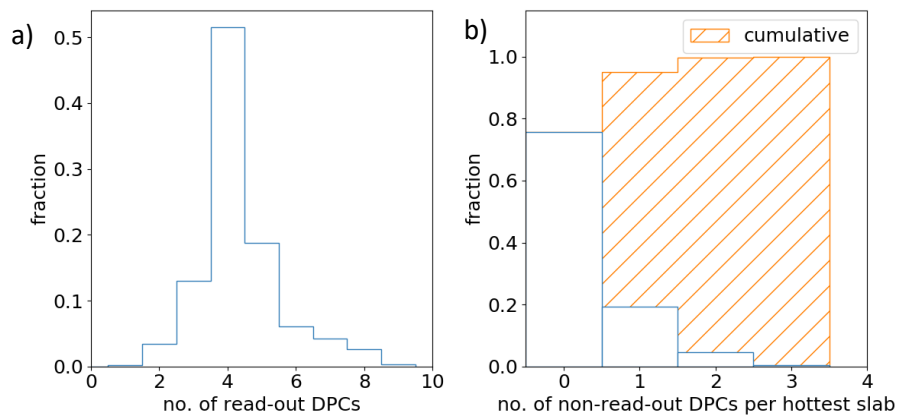

*Figure 4 Readout characteristics of the slab detector. a) Distribution of read-out DPCs per event. b) Distribution and cumulative distribution of non-read-out DPCs (missing DPCs) per hottest slab.*

The distribution of read-out DPCs per event is shown in Figure 4. A clear peak for 4 DPCs corresponding to a fully read-out slab is visible with a mean±SD of 4.35±1.21 DPCs. For the active slab identified by the hottest pixel, a fraction of nearly 75% of events contain data without one or multiple missing DPC hits. Including one missing DPC hit increases the fraction to 95%.

### 4.2. Position Estimation

#### 4.2.1. Planar Positioning

A GTB model of maximum depth 12, learning rate 0.05 and 930 decision trees led to the best-found positioning performance. The results of the planar positioning performance are depicted in Figure 5 and Table 1 for all probed energy windows. An MAE of 1.27 mm was achieved for the narrow energy window as the best result. Both positioning error distribution and cumulative positioning error distribution show the same course for all energy windows. A selection of events closer to the photopeak leads to a better positioning performance. The difference in positioning performance regarding MAE, RMSE, and d90 was between 9.5%-10.5% comparing the wide energy window with no energy window and 20.0%-22.7% comparing the narrow energy window with no energy window, respectively. The d50 shows only minor improvements of 7.1%-8.4% and 15.2%, respectively. The spatial distributions of the MAE and bias vector were found constant over the central area of the detector, with an evident deterioration towards the detector edges. The positioning profiles also confirm this behavior. Separating the detector in two DOI-layers using the corresponding GTB models described in the next Section reveals reduced edge effects for gamma interactions close to the photosensor.

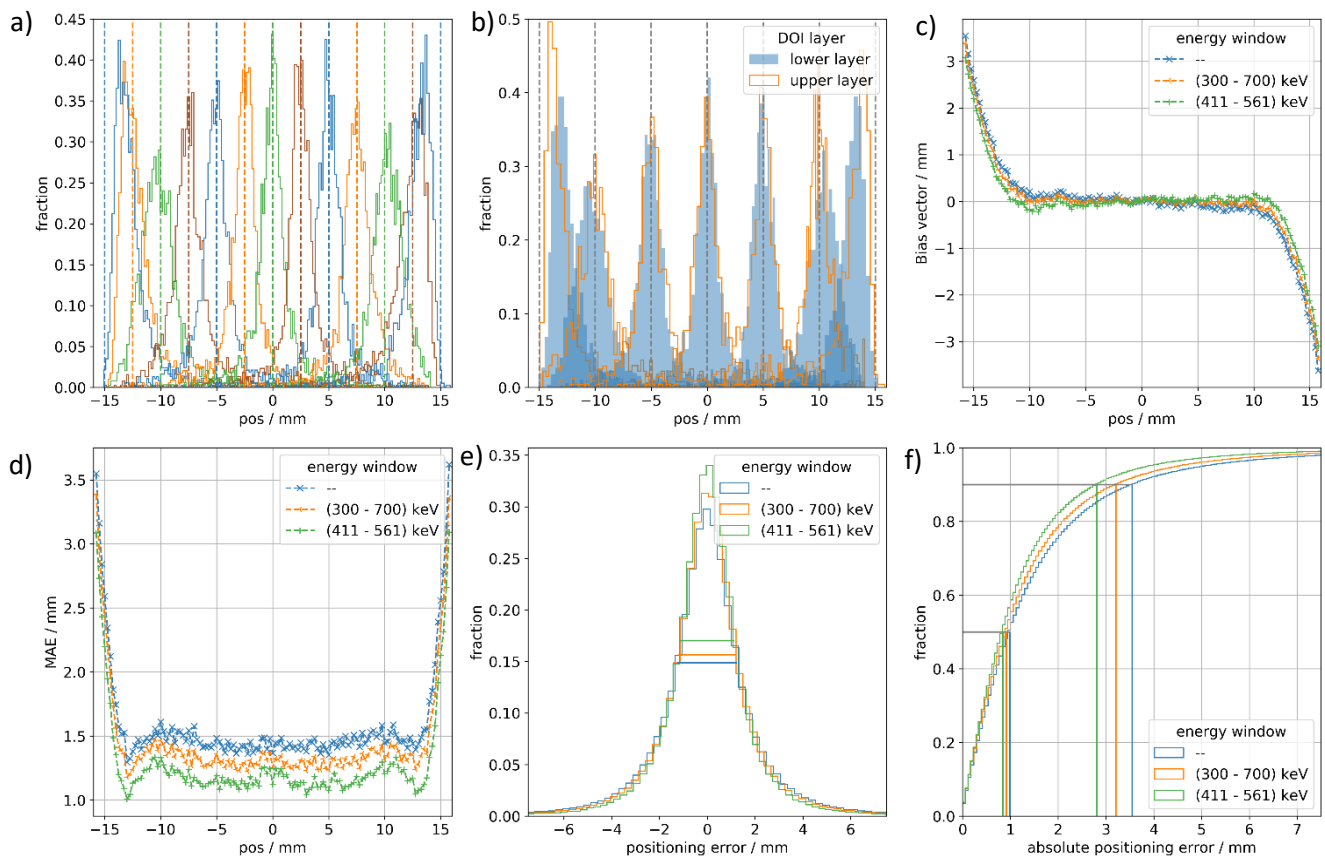

*Figure 5 Evaluation of the planar positioning performance along the monolithic direction. For all figures, the GTB model found according to the high-positioning-performance-optimization was selected (maximum depth 12, learning rate 0.05 and 930 decision trees) and evaluated using the test data set for three energy windows except for panels a) and b), which show all events in an energy window of 300 keV to 700 keV. According to Section 3.1, the test data set contains gamma interactions uniformly distributed along the monolithic and pixelated direction and following an exponential distribution along the DOI-direction according to Lambert-Beer's law. The influence of the finite source diameter was not corrected. a) Positioning profiles exemplarily for irradiation positions ranging from -15 mm to 15 mm with steps of 2.5 mm. The dashed lines display the irradiation position. b) Positioning profiles exemplarily for irradiation positions ranging from -15 mm to 15 mm with steps of 5 mm separated for an upper and lower DOI layer. The DOI layer was identified utilizing the DOI models described in the following Section 4.2.2. The lower layer ranges from 0 mm to 9.5 mm and the upper layer from 9.5 mm to 19 mm with 0 mm representing the photosensor's surface. The dashed lines display the irradiation position. c) Spatial distribution of the Bias Vector along the whole detector. d) Spatial distribution of the MAE along the whole detector. e) Positioning error distribution with indication of the FWHM. f) Cumulative positioning error distribution with indication of d50 and d90.*

*Table 1 Overview of the positioning performance for GTB-based calibrations in both monolithic direction and DOI. The best results are shown without restricting the memory requirement for different energy windows. In the case of no energy window, at least 400 optical photons were required during the cluster process. The performance parameters were not corrected for the finite beam diameter. For positioning along the monolithic direction, a GTB model of maximum depth 12, a learning rate of 0.05 and 930 decision trees was evaluated. For DOI positioning, the same hyperparameters were found except with the number of decision trees to be 990.*

|  | Monolithic Direction | | | DOI | | |
|---|---|---|---|---|---|---|
| Energy window / keV | -- | 300 – 700 | 411 – 561 | -- | 300 – 700 | 411 – 561 |
| MAE / mm | 1.59 | 1.44 | 1.27 | 2.27 | 2.12 | 1.94 |
| RMSE / mm | 2.55 | 2.28 | 1.97 | 3.08 | 2.90 | 2.67 |
| FWHM / mm | 2.50 | 2.27 | 2.29 | 4.08 | 3.92 | 3.51 |
| d50 / mm | 0.99 | 0.92 | 0.84 | 1.66 | 1.55 | 1.40 |
| d90 / mm | 3.55 | 3.21 | 2.81 | 5.11 | 4.78 | 4.35 |
| Bias90 / mm | 1.82 | 1.68 | 1.42 | 3.63 | 3.37 | 2.98 |

### 4.2.2. Depth of Interaction Positioning

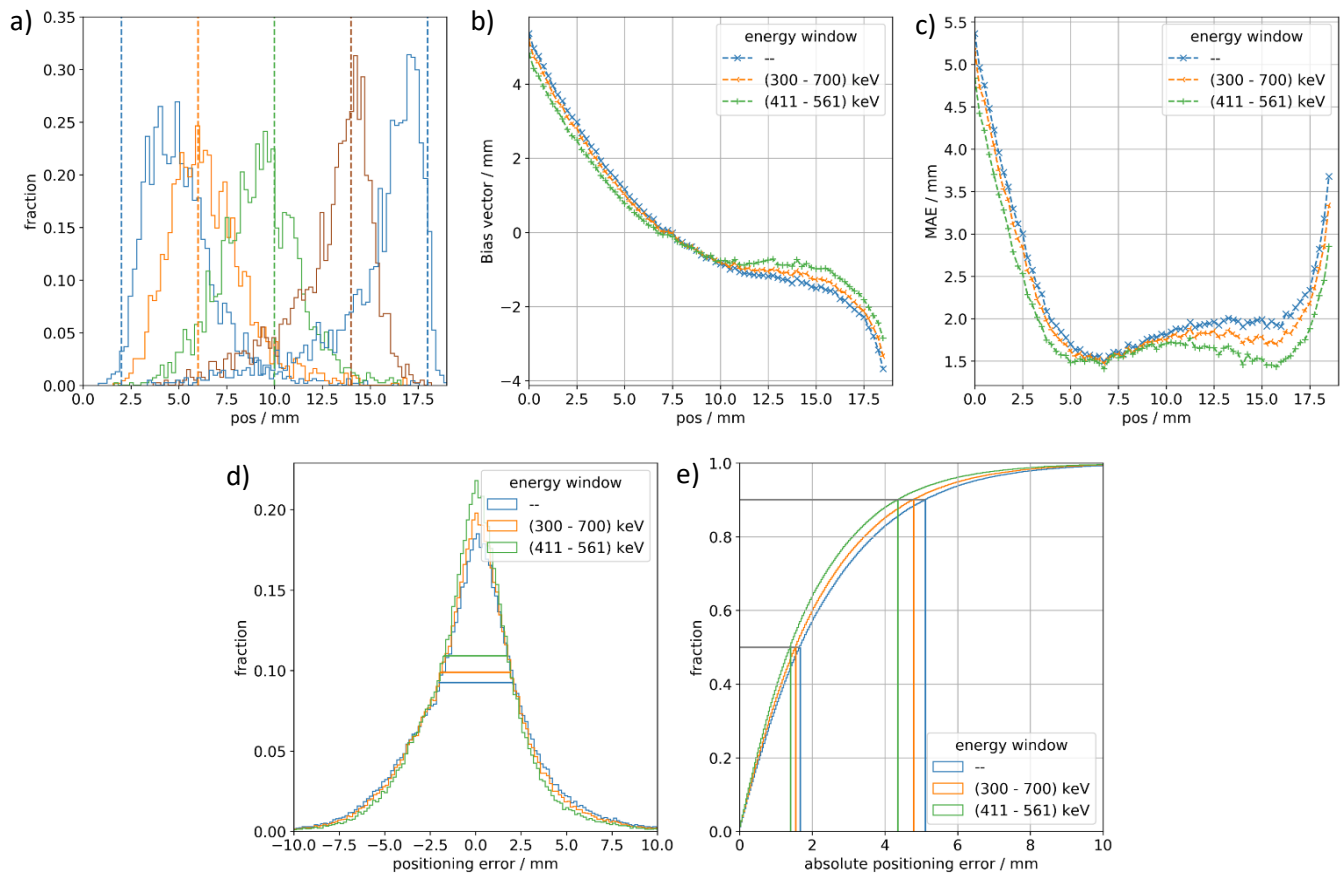

*Figure 6 Evaluation of the DOI positioning performance. For all figures, the GTB model found according to the high-positioning-performance-optimization was selected (maximum depth 12, learning rate 0.05 and 990 decision trees) and evaluated using the test data set for three energy windows except panel a). The influence of the finite source diameter was not corrected. Position 0 mm represents the photosensor's surface. a) Positioning profiles of the test data within an energy range of 300 keV to 700 keV exemplarily for five irradiation positions ranging from 2 mm to 18 mm with steps of 4 mm. The dashed lines display the irradiation position. b) Spatial distribution of the bias vector along the whole detector. c) Spatial distribution of the MAE along the whole detector. d) Positioning error distribution with indication of the FWHM. e) Cumulative positioning error distribution with indication of d50 and d90.*

A GTB model of maximum depth 12, learning rate 0.05, and 990 decision trees led to the best-found DOI positioning performance. The best MAE was spotted to be 1.94 mm for the narrow energy window. The positioning profiles, spatial distributions of bias vector and MAE as well as positioning error distributions are depicted in Figure 6; the averaged performance parameters are shown in Table 1. The positioning error distribution indicates a slight asymmetrical behavior close to the tail regions, which is represented in the bias vector and MAE distribution as well. The MAE performance's deterioration is evident for up to 5 mm from the photosensor surface and 2.5 mm from the scintillator surface. More narrow energy windows lead to a more homogenous positioning performance. The difference in positioning performance regarding all tested parameters varied between 5.8%-6.6% comparing the wide energy window with no energy window and 13.3%-15.6% comparing the narrow energy window with no energy window, respectively.

### 4.2.3. Optimizations on Computational Requirements

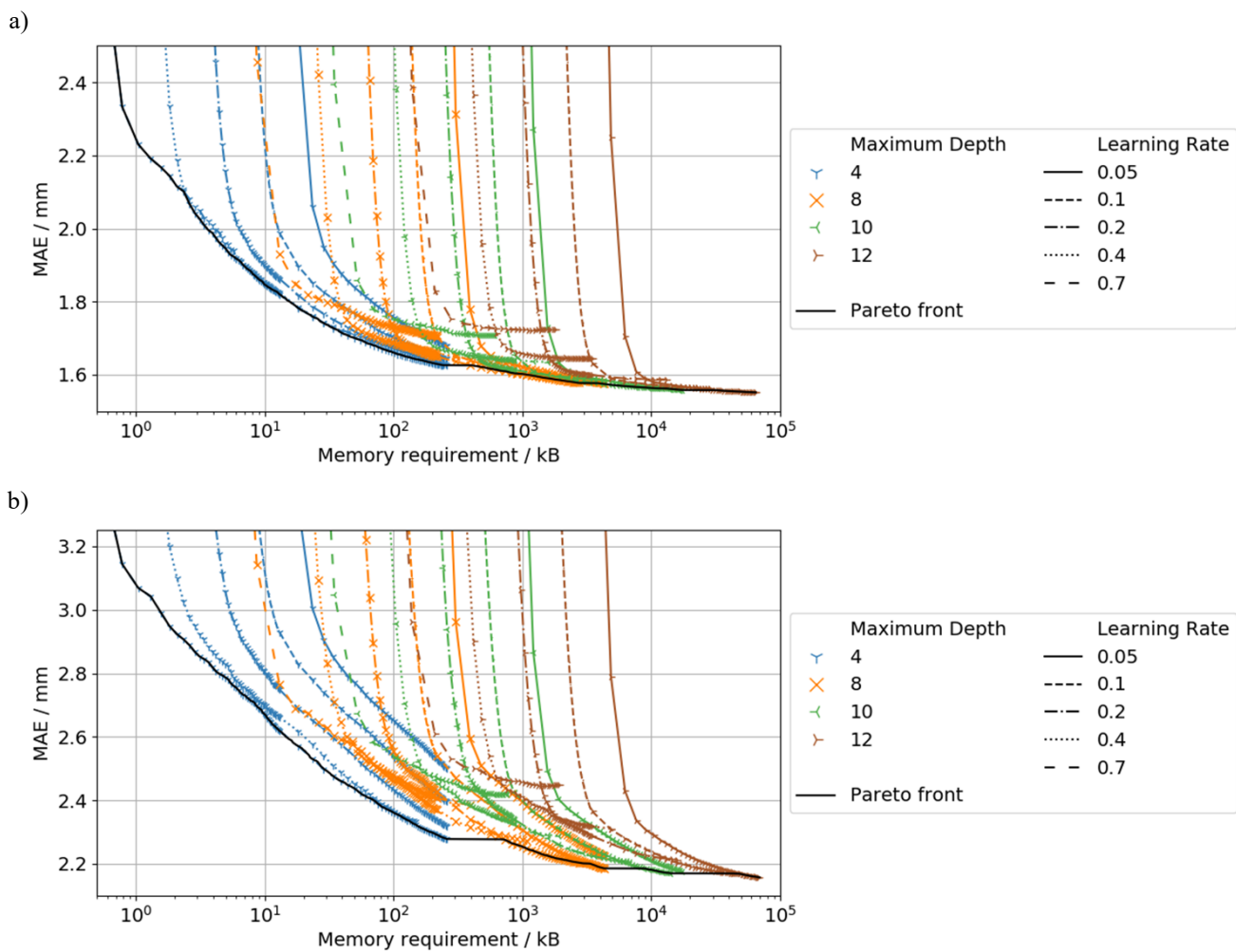

*Figure 7 Pareto Frontier of positioning performance in terms of MAE versus memory requirement for a) planar positioning and b) DOI positioning. No energy filter was applied. The graphs display the course of the MAE for ensembles of varying number of decision trees for fixed maximum depths (color-coded and markers) and learning rates (coded as line style). GTB models allowing a higher maximum depth achieve a better positioning performance at the cost of an increased memory requirement. High learning rates favor reduced memory requirements at the cost of positioning performance for a given maximum depth. The Pareto Frontier enables selection of the best compromise for both shown parameters.*

The Pareto frontier is shown in Figure 7 for both planar and DOI positioning, demonstrating the versatility of GTB-based positioning models. The previously defined hyperparameters lead to a maximum memory requirement of approx. 25 MB using a maximum depth of 12 and 1000 decision trees. However, the memory requirement can be shrunk by several orders of magnitude down to a few kB for both positioning directions with a reasonable loss in positioning performance. In general, GTB models of higher maximum depth require more memory resources while achieving typically better positioning performance. High learning rates favor less complex GTB models at the cost of the best possible positioning performance for a constant maximum depth and tend to overfitting. The Pareto frontier indicates an advantage of using models of small maximum depth in case of strict restrictions on the memory requirement.

### 4.3. Energy Calibration and Estimation

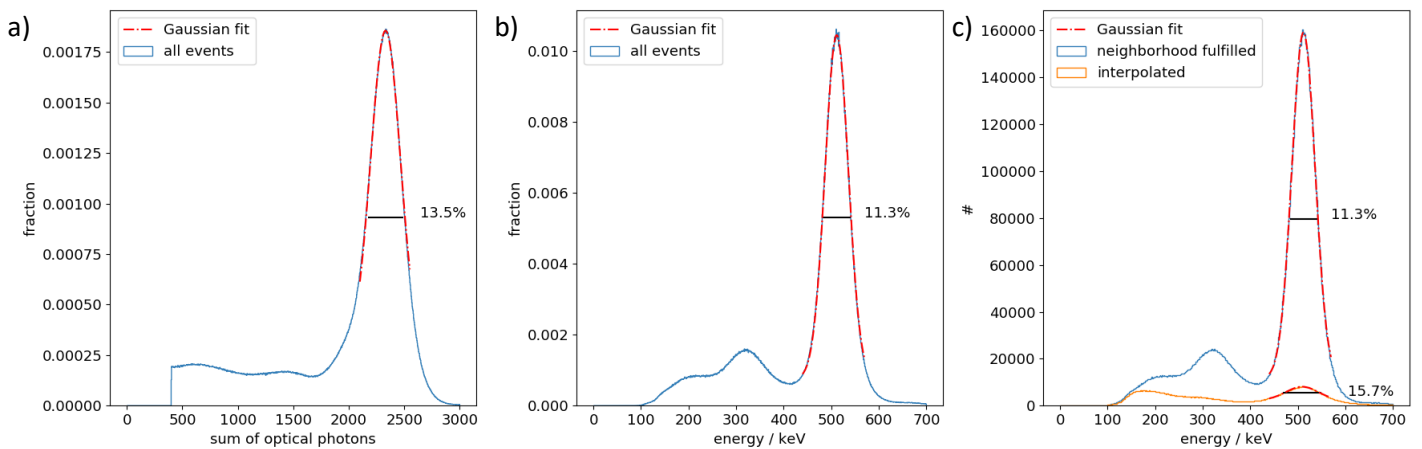

*Figure 8 Photon-count and energy distribution. The dashed line represents a Gaussian fit performed in the indicated fitting range and the FWHM is indicated in the plot. a) Distribution of the detected optical photon of the whole slab detector with an estimated width of 13.5%. No gamma interactions with less than 400 optical photons were included in the data processing. b) Normalized energy distribution for all measured events (dE/E = 11.3%). c) Separated energy distributions for events fulfilling the chosen neighborhood criterion DiesFN (dE/E 11.3%) and interpolated ones (dE/E 15.7%). The distributions are not normalized to represent the proportions during real measurement conditions.*

The uncalibrated distribution of the sum of optical photons shows a clear 511 keV-photopeak with a width of 13.5% FWHM (see Figure 8). The energy calibration correcting for 3D-dependent light detection variation and missing DPC hits achieves an energy resolution dE/E of 11.3%. A fraction of 13% of all events does not fulfill the chosen neighborhood criterion DiesFN, while 25% of all events are missing at least one DPC hit. This is caused by the definition of the DiesFN, which is less strict in requiring two or three available DPC hits in contrast to all DPC hits to be present (compare its definition in Section 3.3). Those events show a deteriorated energy resolution of dE/E 15.7%.

### 4.4. Timing Performance

The general timing performance was once evaluated performing only the analytical timing calibration, and once performing additionally the machine learning-based timing calibration.

#### 4.4.1. Performance After Analytical Calibration

Regarding the analytical calibration, two measurement configurations were used, consisting of a slab and a one-to-one coupled detector or two one-to-one coupled detectors, allowing a better comparison between the two crystal architectures. As shown in Table 2, a slab setup is expected to show a timing performance comparable to a one-to-one coupled setup.

*Table 2 Obtained time resolutions after analytical timing calibration for different detector configurations and energy windows. Filter F ensures that the first timestamp originates from the hottest DPC sensor, the hottest pixel is on the hottest DPC sensor, and the active slab is completely read out.*

|  | CRT [ps] | | |
| --- | --- | --- | --- |
| Detector combination | 1-to-1 vs. 1-to-1 | slabs vs. 1-to-1 | slabs vs. slabs |
| $E \in [300, 700]$ keV | 234 | 237 | 240 |
| $E \in [411, 561]$ keV | 228 | 231 | 234 |
| $E \in [411, 561]$ keV $\wedge$ F | 224 | 223 | 222 |

### *4.4.2. Performance After Analytical & Machine Learning-Based Calibration*

The results obtained with the mixed setup (slab vs. one-to-one) after performing analytical and machine learning-based calibration are compared to performing only the first two analytical calibration stages using the same crystal configuration. The best performing GTB model was found with maximum depth 20, 22 decision trees and learning rate of 0.1. As shown in Table 3, the machine learning-based timing calibration strongly improves the achievable time resolutions, such that CRT values below 200 ps are possible for a setup comprising at least one slab detector. Applying additional quality cuts on the light distribution (filter F as defined before), further enhances the CRT to 186 ps at the cost of dismissing 50% of the measured coincidences.

*Table 3 Comparison between performing only the (first two stages of the) analytical calibration and the machine learning-based calibration.*

| Filter | CRT [ps] | |
|---|---|---|
| | analyt. calibration | analyt. & ML calibration |
| $E \in [300, 700]$ keV | 237 | 208 |
| $E \in [411, 561]$ keV | 228 | 198 |
| $E \in [411, 561]$ keV $\wedge$ F | 224 | 186 |

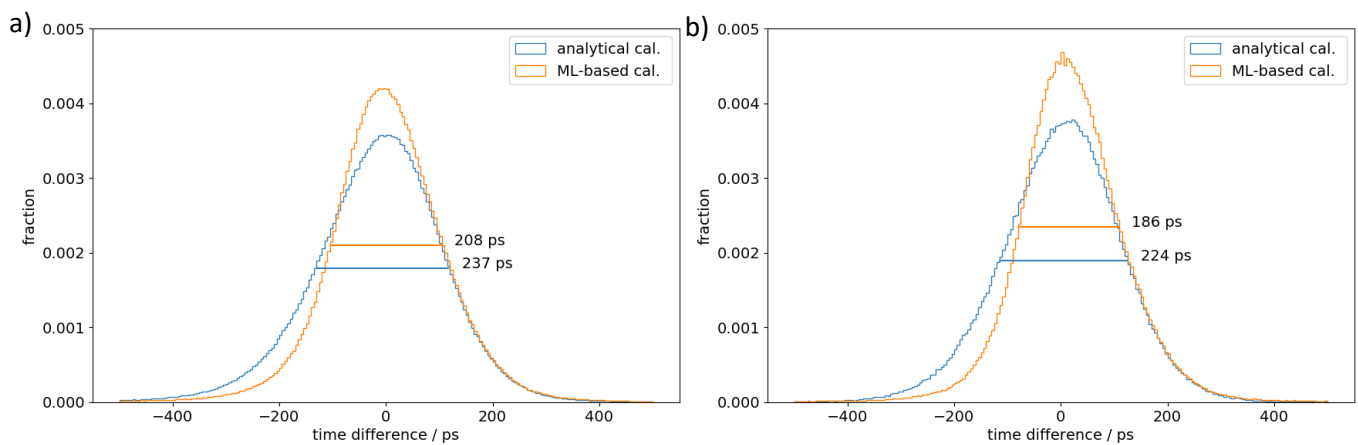

*Figure 9 Time difference histograms for the analytical and machine-learning-based timing calibration schemes. The machine-learning-based timing calibration significantly improves the CRT. The GTB model mainly effects the left tail of the distribution which is dominated by timestamps of the slab detector. a) Wide energy window with no further quality cuts applied. b) Narrow energy window with additional filter F operating on the light distribution.*

## 5. Discussion

As listed in the introduction, several groups demonstrated attractive performance parameters for semi-monolithic crystals with a pre-clinical focus. Overall, the presented characteristics make them a suitable candidate to explore including DOI information into clinical PET systems, focusing on a good CRT.

Compared to monolithic scintillators known from previous studies, we have found a significantly higher density of optical photons registered by the photosensor. Furthermore, at least three DPCs of the current slab sent hits for the majority of events. This significantly reduces the challenges of missing DPC hits during the data processing. In contrast, we registered only 9 out of 16 DPC hits for a monolith of comparable size in previous work. Nevertheless, all presented calibration steps and algorithms include events independent of the number of missing DPCs to maintain the maximum achievable sensitivity of the detector. The calibration and processing techniques are directly transferable to other readout and sensor technologies, e.g., analog SiPMs. The issue of missing information of single readout channels or regions, for example, is shared with other solutions digitalizing single readout channels, e.g., the TOFPET2 ASIC[74,75]. As the dead time is the main reason for missing information, it can serve as a measure for comparing both digitization approaches. The DPC undergoes a dead time of up to 80 ns for both validated and non-validated triggers recharging the single SPADs and requires 680 ns for readout.[76] This compares to a minimum dead time of 100 ns for the "triggered" state and 25 ns for the "validated" state of the TOFPET2 ASIC.[77] Therefore, we expect a lower rate of missing information for the TOFPET2 ASIC, while the distribution highly depends on the detector design, photosensor, its dark current, as well as specific settings of the ASIC and are thus hard to predict.

We set the current measurement DPC temperature to around 5 °C. Although challenging for implementation on full system level, this temperature can be realistically reached using liquid cooling systems, as demonstrated for a high grade of integration in spatial demanding PET/MRI applications.[78] For higher measurement temperatures, the dark count rate of the detectors is expected to increase, resulting in many non-validated triggers and thus increased dead-time. To preserve the detector sensitivity, one would need to relax the DPC trigger scheme requiring more optical photons, which typically deteriorates the timing resolution.[61] From experience in our lab, the detector could be operated up to room temperature, sacrificing most of the timing capabilities, which requires verification in future work.

The fan beam collimator enables a time-efficient procedure for positioning calibration along both monolithic- and DOI-direction. Inherent to the principle of semi-monolithic detectors, only one planar positioning calibration needs to be conducted. Model training of a single GTB model can be done in the order of minutes with a standard computer. Taking the point, that the model training needs only to be done when calibrating or re-calibrating the detector, this time seems very reasonable transferring the method to many detectors. The established GTB-based positioning models achieved a decent positioning performance of 1.44 mm MAE in planar and 2.12 mm MAE in DOI direction (wide energy window), especially considering the presumed clinical application. Future research could detailly test and optimize the influence of the number of missing DPC hits per gamma interaction on the positioning performance.

The memory requirement optimization enables a reduction of the memory requirement of orders of magnitudes down to the kB-range and allows to chose highly optimized positioning models for given computational resources. The Pareto-frontier concept provides an easy and efficient selection process. As an advantage to our previous work, this method is independent of the current studied detector, while previously dedicated gradient conditions (e.g., improvement of MAE per added decision tree) were established to select the most memory-efficient GTB models.

The feasibility of applying GTB-models in software-based online processing was recently demonstrated and characterized for the Hyperion-II$^D$ platform.[44,79] For the Hyperion II$^D$ setup, we achieved processing throughputs of 9.5 GB/s corresponding to approximately $2 \cdot 10^6$ events/s or a processing time of 500 ns/event. For the successor of the Hyperion II$^D$ platform, the inference of GTB models was demonstrated in firmware with processing throughputs of up to $4.55 \cdot 10^6$ events/s or a processing time of approximately 200 ns/event for every individual detector.[71,72] These results validate that the established calibration routines can be transferred to full PET systems. The current study estimated DOI as a continuous parameter using a GTB regression model to characterize the full DOI capabilities. For clinical PET systems, a multi-layer DOI encoding into 2-4 layers may be sufficient to significantly reduce the radial astigmatism effects.[80,81] Thus, the precision of the DOI estimation and corresponding computational requirements could be further relaxed if required.

The energy calibration relies on a flood irradiation without a collimator setup and can be conducted in an assembled PET system as well. The slab-detector showed an attractive energy resolution of 11.3%. The selection of a region with mostly stable readout conditions, i.e., no missing DPC hits, is favorable during the energy calibration. The interpolation method compensates for single missing DPC hits, though, leads to a deterioration of the energy resolution for this event type. More advanced interpolation methods may improve the energy estimation for those events.

We established both an analytical and machine-learning-based timing calibration. The analytical timing calibration requires data from a few source positions to establish an overdetermined matrix system based on feasible channel combinations of the photosensor pair. This procedure could be easily translated to assembled PET systems. The principle of determining specific electrical and optical time skews was already described in the literature.[20,21,61] Here, we extended this principle by assuming the channel pairs to be abstract objects, e.g., representing simple crystal volumes or energy bins. The calibration procedure reveals simple time offsets, which can be applied computationally efficiently on the level of PET systems. We achieved a good CRT of 240 ps for a pair of slab detectors using the wide energy window without any further quality cuts. This CRT is comparable to or better than currently available clinical PET scanners.

Furthermore, we implemented a novel machine-learning-based timing calibration utilizing GTB, which is already established for position estimation. The GTB model training requires training data with all possible physical time differences. Thus, a source or an array of sources needs to be translated between both detectors or through the full scanner volume for an assembled PET ring. Further research will tackle this point to reduce the calibration effort. Nonetheless, the GTB-based timing calibration greatly improves the CRT and demonstrates the full potential of the slab detector concept. For the wide energy window and without any further quality cuts, we achieved 209 ps CRT and 186 ps with a narrow energy window and filtering on the light distribution. The direct comparison of the time difference histograms unveils the main improvement emerging from the tail dominated by delayed timestamps of the slab detector. The GTB model utilizes additional information not directly accessible in the matrix-formulation of time skews. Besides GTB, also other algorithms may be suitable for predicting the time difference of gamma interactions. However, we again chose GTB as this algorithm is utilizable for online software processing.

Overall, the semi-monolithic detector provides attractive positioning, energy, and time resolution performance characteristics. Especially, a good CRT can be maintained while introducing DOI capabilities to the detector. Thus, we see the detector concept as suitable for clinical PET scanners. The presented calibration aims at time-efficient methods which apply to the scale of full PET systems.

## 6. Conclusion

We established the calibration process for a semi-monolithic detector tackling the specific demands of clinical PET applications. The detector achieves an attractive positioning, energy, and time resolution performance. We found an MAE of 1.44 mm for planar and 2.12 mm for DOI positioning in a wide energy window of [300, 700] keV for positioning models based on a fan beam collimator calibration and GTB. The algorithm allows a trade-off between positioning performance and computational requirements and allows selection of those GTB models suitable for specific data processing topologies. The energy resolution was shown to be 11.3%. We achieved a CRT of 240 ps for the wide energy window for an analytical timing calibration. This value was improved by applying a GTB timing estimation to 208 ps for the wide energy window, demonstrating additional potential regarding CRT optimization when utilizing all available detector information. All results were achieved for data processing techniques taking all registered gamma interactions into account and thus preserving the full sensitivity of the detector. Therefore, we see the detector concept as suitable for clinical PET scanners. Further research may shape the detector characteristics by adjusting the detector wrapping. The fan-beam-collimator concept can be enhanced and translated to an in-system calibration. The GTB-based timing estimation proves great potential for further application and translation to other detector topologies.

**Acknowledgments**

We thank Axel Honné and Adalbert Mazur - representative for the whole team - of the Scientific Workshop of the University Hospital Aachen for manufacturing all mechanical components utilized in this study.

**Conflict of Interest**

The authors have no relevant conflicts of interest to disclose.

**Data Availability Statement**

The data that support the findings of this study are available from the corresponding author upon reasonable request.

**Appendix I: Mathmatical description of the analytical timing calibration**

This appendix describes the analytical timing calibration presented in Section 3.4.1 and illustrated in Figure 3 in more detail. Already introduced abbreviations are valid for this section as well. Each of the two facing detectors $a$ and $b$ is separated into abstract objects, which can be represented by physical readout channels (like pixels) but also by theoretical constructs (like scintillator volumes). Assuming channels as example for the following expressions, one obtains a set of $n_a$ channels for detector $a$ and a set of $n_b$ channels for detector $b$ with $c_i$ denoting a channel,

$$\mathcal{C}^a = \{c_1, \ldots, c_{n_a}\},$$
$$\mathcal{C}^b = \{c_{n_a+1}, \ldots, c_{n_a+n_b}\}.$$

Two mapping functions $\phi_a, \phi_b$ exist, which uniquely assign each coincidence $i$ to a corresponding channel pair of detector $a$ and $b$,

$$\exists \phi_a, \phi_b : \begin{cases} \phi_a(i) \mapsto c_k, \text{ for } c_k \in \mathcal{C}^a \\ \phi_b(i) \mapsto c_l, \text{ for } c_l \in \mathcal{C}^b \end{cases}$$

For each channel combination $(i,j)$ with $c_i \in \mathcal{C}_a$ and $c_j \in \mathcal{C}_b$ a time difference set $\mathcal{D}_{kl}$ can be defined comprising $n_{kl}$ timestamp differences,

$$\mathcal{D}_{kl} \equiv \{(t_k - t_l)_i\}_{i=1}^{n_{kl}},$$

with $t_{f,g}$ denoting the timestamp of coincidence $g$, that has been assigned to channel $c_f$. During the formulation of the matrix equation, one tries to produce equations like,

$$mean(\mathcal{D}_{kl}) = mean(\{t_k - t_l\}_{i=1}^{n_{kl}}) \stackrel{!}{=} \overline{\Delta t_{kl}} = c_k - c_l + t_{\text{off}}^{global}$$

where $t_{\text{off}}^{global}$ accounts for a possible global off-centred source position and the mean of the time difference set is estimated by fitting a Gaussian function to the corresponding distribution. In case of simultaneously calibrating multiple detectors (n>2) with multiple sources, additional offsets need to be introduced accounting for induced time shifts. Therefore, a channel vector $\vec{c}$,

$$\vec{c} = \begin{pmatrix} c_1 \ldots c_k \ldots c_{n_a} & c_{n_a+1} \ldots c_l \ldots c_{n_a+n_b} \end{pmatrix}^T,$$

as well as a mean time difference vector $\overrightarrow{\Delta t}$,

$$\overrightarrow{\Delta t} = \left(\overline{\Delta t}_{(1,n_a+1)} \ldots \overline{\Delta t}_{kl} \ldots \overline{\Delta t}_{(n_a,n_a+n_b)}\right),$$

are defined. Together, with an indexing matrix $\underline{M}$ the final matrix equation is given to be

$$\underline{M} \cdot \vec{c} = \overrightarrow{\Delta t}.$$

As global constraint, two equations are added to the matrix $\underline{M}$ ensuring the skews sum up to zero for each detector. The analytical corrections $\overrightarrow{c_{min}}$ were subsequently obtained by minimizing the Euclidean-2-norm,

$$\overrightarrow{c_{min}} = \underset{\vec{c}}{\text{argmin}} \, ||\overrightarrow{\Delta t} - \underline{M} \cdot \vec{c}||_2.$$